\documentclass{article}
\usepackage{subcaption}
\usepackage{enumitem} 
\usepackage{authblk} 

\usepackage{graphicx} 
\usepackage{amsmath} 
\usepackage{amssymb} 
\usepackage{amsthm} 
\usepackage{booktabs}
\usepackage{hyperref} 
\usepackage{enumitem} 
\newtheorem{theorem}{Theorem}

\title{Emergent Collective Memory in Decentralized Multi-Agent AI Systems}

\author{Khushiyant}
\affil{University of Freiburg}

\begin{document}

\maketitle

\begin{abstract}
We demonstrate how collective memory emerges in decentralized multi-agent systems through the interplay between individual agent memory and environmental trace communication. Our agents maintain internal memory states while depositing persistent environmental traces, creating a spatially distributed collective memory without centralized control. 

Comprehensive validation across five environmental conditions (20×20 to 50×50 grids, 5-20 agents, 50 runs per configuration) reveals a critical asymmetry: individual memory alone provides 68.7\% performance improvement over no-memory baselines (1563.87 vs 927.23, p<0.001), while environmental traces without memory fail completely. This demonstrates that memory functions independently but traces require cognitive infrastructure for interpretation.

Systematic density-sweep experiments ($\rho \in [0.049, 0.300]$, up to 625 agents) validate our theoretical phase transition prediction. On realistic large grids (30×30, 50×50), stigmergic coordination dominates above $\rho \approx 0.20$, with traces outperforming memory by 36-41\% on composite metrics despite lower food efficiency. The experimental crossover confirms the predicted critical density $\rho_c = 0.230$ within 13\% error. 

\end{abstract}

\section{Introduction}

Coordinating multiple autonomous agents in complex and dynamic environments remains a fundamental challenge in the fields of swarm robotics, distributed artificial intelligence, and agent-based modeling \cite{van2008multi}. In such systems, efficient exploration, resource acquisition, and adaptation hinges on how agents perceive, store, and share information about their surroundings. While direct communication protocols can facilitate rapid information transfer, they often require structured networks and incur scalability costs. In contrast, many natural systems, such as ant colonies and termite swarms, achieve remarkable coordination through \emph{stigmergy}---an indirect communication paradigm in which agents influence one another via persistent environmental modifications, such as pheromone trails \cite{di2022evolution}.

Recent work has advanced decentralized and adaptive multi-agent collaboration strategies systems \cite{tang2024decentralizedlifelongadaptivemultiagentcollaborative}, and lifelong shared-memory systems for cooperative pathfinding \cite{sagirova2025srmtsharedmemorymultiagent}, and decentralized cooperative belief modeling for dynamic tasks \cite{zhai2023dynamic}. These approaches highlight the growing interest in combining local decision-making with persistent or shared cognitive structures for scalable coordination.

Inspired by these biological precedents, our work investigates how environmental traces can serve as a scalable substrate for \emph{emergent collective memory} in decentralized multi-agent systems. We design agents that integrate individual, category-specific memories with the ability to leave and interpret environmental traces encoding information about food sources, hazards, social encounters, and exploration patterns. This combination enables agents to act upon both personal experience and a shared, spatially distributed memory that emerges from group interaction with the environment.

\section{Related Works}

The study of decentralized multi-agent systems has long explored mechanisms for coordination, communication, and memory to enable robust collective behavior. A well-established paradigm is \emph{stigmergy}, wherein agents communicate indirectly through modifications of the environment, such as pheromone trails in biological systems, enabling scalable coordination without centralized control \cite{bonabeau1999swarm,boldini2024stigmergy,de2020multi}. Recent applications demonstrate stigmergic coordination in robot swarms for automatic behavior design \cite{salman2024automatic} and radioactive environment exploration \cite{ardiny2024enhancing}. Our work builds on this principle by formalizing a multi-category environmental trace system that encodes task-relevant information to support emergent collective memory.

Recent frameworks such as decentralized lifelong-adaptive collaborative learning \cite{tang2024decentralizedlifelongadaptivemultiagentcollaborative} and shared memory architectures for multi-agent pathfinding \cite{sagirova2025srmtsharedmemorymultiagent} demonstrate the potential of persistent, distributed knowledge stores. Adaptive partner modeling for cooperation in dynamic environments \cite{xu2024decentralized} has further shown how modeling other agents' behavior improves coordination robustness. Dynamic belief-based methods for decentralized cooperation \cite{zhai2023dynamic} also contribute to scalable joint decision-making.

Experimental testbeds like GenGrid \cite{kedia2021gengrid} have provided flexible simulation environments for swarm robotics, supporting research into stigmergic communication and environmental interactions. Meanwhile, large language model-driven frameworks such as LLM-powered collective tuning \cite{chen2025fostering} and decentralized evolutionary coordination \cite{yang2025agentnet} extend multi-agent collaboration into high-level cognitive domains.

Beyond engineering, stigmergy has been analyzed rigorously from a control-theoretic perspective \cite{boldini2024stigmergy} and adapted through virtual pheromone systems for robot coordination \cite{de2020multi}. Emergent exploration capabilities have been demonstrated by decentralized meta-reinforcement learning systems \cite{bornemann2023emergence}, showing how indirect coordination can yield open-ended collective behaviors.

In sum, our proposed framework advances the state-of-the-art by combining biologically inspired stigmergic communication with multi-categorical memory organization and consensus weighting, integrating ideas from both recent AI research \cite{chen2025fostering,yang2025agentnet} and theoretical stigmergy modeling \cite{boldini2024stigmergy}.

\section{Methodology}

\subsection{Agent-Based Model Architecture}

We designed a decentralized multi-agent framework to investigate emergent collective memory and coordination. The implementation uses the Mesa agent-based modeling framework \cite{masad2015mesa}. Agents operate on a discrete two-dimensional grid, interacting solely through local perception and persistent environmental traces, with no centralized controller. Collective intelligence thus arises from individual memory dynamics and indirect stigmergic communication.

Each agent maintains a personal memory organized into four categories: food, danger, social, and exploration memories, denoted as
\begin{equation}
\mathcal{M}_a = \{ \mathcal{M}_a^{\text{food}}, \mathcal{M}_a^{\text{danger}}, \mathcal{M}_a^{\text{social}}, \mathcal{M}_a^{\text{exploration}} \}
\end{equation}

Each memory entry \( m \in \mathcal{M}_a^{c} \) encodes a tuple \( (\mathbf{p}, t_m, s_m) \), where spatial location \( \mathbf{p} \in \mathbb{Z}^2 \), acquisition time step \( t_m \), and confidence strength \( s_m \) together represent agent experience.

Memory strengths decay exponentially over time with category-dependent decay rates \( \delta_c \):
\begin{align}
s_m(t + \Delta t) &= s_m(t) \times \delta_c^{\Delta t}, \\
\delta_{\text{food}} = 0.985, \quad 
\delta_{\text{danger}} = 0.998, \\
\delta_{\text{social}} = 0.95, \quad 
\delta_{\text{exploration}} = 0.97
\end{align}
These rates reflect memory importance: danger memories ($\delta$ = 0.998) persist longest since threats remain relevant, while social memories ($\delta$ = 0.95) decay fastest as agent configurations change rapidly. Food memories ($\delta$ = 0.985) balance persistence with adaptability to resource depletion.

Memories falling below a threshold \( s_{\mathrm{thresh}} = 0.2 \) or exceeding maximal age are pruned to limit memory size, with a default capacity of 50 entries per agent.

Agents possess energy levels \( E_a(t) \in [0, E_{\max}] \) with \( E_{\max} = 150 \), initialized at 100. Energy expenditures include movement cost \( E_{\mathrm{move}} = 1 \), trace generation cost \( E_{\mathrm{trace}} = 2 \), and baseline metabolism, balanced against energy gains from resting and food consumption:
\begin{equation}
E_a(t+1) = E_a(t) - E_{\mathrm{move}} \cdot m_t - E_{\mathrm{trace}} \cdot l_t - E_{\mathrm{base}} + E_{\mathrm{food}} + E_{\mathrm{regen}}
\end{equation}
where \( m_t, l_t \in \{0,1\} \) indicate movement and trace leaving events at time \( t \).

Behavioral characteristics such as exploration tendency, social learning rate, memory trust, and cooperation tendencies are drawn from predefined distributions to induce heterogeneous agent profiles.

Agents cycle dynamically among task states—exploring, foraging, returning, resting—based on internal energy and environmental cues. Movement decisions are computed by scoring candidate neighboring positions \( \mathbf{p} \in \mathcal{N}(\mathbf{p}_a(t)) \) via weighted sums:
\begin{equation}
S_a(\mathbf{p}, t) = R_a(\mathbf{p}, t) + W_{\mathrm{task}} T_a(\mathbf{p}, t) + W_{\mathrm{mem}} M_a(\mathbf{p}, t) + S_a^{\mathrm{social}}(\mathbf{p}, t) - D_a(\mathbf{p}, t)
\end{equation}
where \( R_a \) is random noise, \( T_a \) task desirability, \( M_a \) memory consensus score, \( S_a^{\mathrm{social}} \) social attraction, and \( D_a \) danger penalty.

Weights are set as
\begin{equation}
W_{\mathrm{task}} =
\begin{cases}
10 & \text{foraging}\\
8  & \text{exploring}
\end{cases}, \quad W_{\mathrm{mem}}=15
\end{equation}
prioritizing memory consensus ($W_{\mathrm{mem}} = 15$) over immediate task desirability, reflecting that collective knowledge should guide individual actions. Foraging receives slightly higher weight than exploration (10 vs 8) to bias energy-depleted agents toward known food sources rather than risky exploration. Agents select the position \( \mathbf{p}^* \) maximizing \( S_a(\mathbf{p}, t) \).

\subsection{Environmental Trace System}

Agents communicate indirectly by depositing \emph{environmental traces}, persistent markers encoding four categories: food, danger, social, and exploration. Trace strengths \( \sigma_c \in [0.7, 1.0] \) scale with agent energy and event significance, and decay exponentially with category-specific rates \( \delta_c' \):
\begin{equation}
\sigma_c(t+1) = \sigma_c(t) \times \delta_c'
\end{equation}

Trace deposition is conditional upon agent state and local context. For example, food traces are left if an agent is carrying food and energy is above 50, danger traces if energy is below 20, social traces in the presence of at least two nearby group members within radius 2, and exploration traces are generated probabilistically at 30\% to reduce noise.

At location \( \mathbf{p} \), collective consensus strength \( C_c(\mathbf{p}, t) \) for trace type \( c \) is computed as
\begin{equation}
C_c(\mathbf{p}, t) =
\begin{cases}
\min \left( 2.0, 1.0 + \alpha (N_c(\mathbf{p}, t) - 1) \right), & N_c(\mathbf{p}, t) \geq 2 \\
0.8 \times \overline{\sigma}_c(\mathbf{p}, t), & \text{otherwise}
\end{cases}
\end{equation}
where \( N_c \) is the count of distinct trace-leaving agents, \( \alpha=0.3 \) an amplification factor, and \( \overline{\sigma}_c \) the average trace strength.

Agents update personal memories using consensus-weighted trace inputs as
\begin{equation}
s'_m = s_m \times C_c(\mathbf{p}, t) \times \beta_{a,c}
\end{equation}
with agent-specific social learning weights \( \beta_{a,c} \) modulating sensitivity to trace consensus.

\section{Mathematical Framework}

\subsection{Overview: Modeling Collective Memory Emergence}

To understand \emph{when} and \emph{why} collective intelligence emerges, we develop a mathematical theory capturing the feedback loop between individual learning and environmental communication. The key question: \emph{How many agents are needed before individual memories coalesce into collective coordination?} Our framework predicts a critical density threshold—analogous to phase transitions in physics—below which agents operate independently, and above which they exhibit synchronized group behavior.

We model the system as interactions between three layers: (1) \textbf{individual agent memories} that store and decay personal experiences, (2) \textbf{environmental traces} that agents deposit and read, creating shared information, and (3) \textbf{the coupling between them} that amplifies or suppresses collective effects. This formalism allows us to derive testable predictions about critical densities and validate them experimentally.

\subsection{Formal Definitions}

We consider a multi-agent memory system defined as a tuple $\mathcal{S} = (G, A, M, T, f, g)$, where $G = \{(i,j) : 1 \leq i,j \leq n\}$ represents a discrete $n \times n$ grid environment, $A = \{a_1, a_2, \ldots, a_k\}$ is the set of $k$ agents, and $M_i(t)$ denotes the personal memory state of agent $i$ at time $t$. The environmental trace intensity at spatial position $(x,y)$ and time $t$ is represented by $T(x,y,t)$. Agent decisions are governed by a function $f$ mapping memories and traces to actions, while $g$ updates both agent memories and the environment based on performed actions.

\subsection{Individual Memory Dynamics}

Each agent's memory evolves continuously according to the balance of information acquisition, forgetting, and social influence. Formally, the time derivative of agent $i$'s memory state is given by the equation
\begin{equation}
\frac{dM_i(t)}{dt} = \alpha \cdot \text{Acquisition}_i(t) - \beta \cdot \text{Decay}(M_i(t)) + \gamma \cdot \text{SocialInfluence}_i(t),
\end{equation}
where acquisition aggregates spatial information from the agent’s vicinity weighted by environmental signals,
\begin{equation}
\text{Acquisition}_i(t) = \sum_{(x,y) \in \mathrm{Vicinity}(a_i)} I(x,y,t) \cdot \phi(x,y),
\end{equation}
decay models exponential forgetting of stored memories,
\begin{equation}
\text{Decay}(M_i(t)) = \sum_j \lambda_j m_{i,j}(t) e^{-\delta \tau_j},
\end{equation}
and social influence captures shared memory effects from neighboring agents weighted by coefficients $w_{ij}$,
\begin{equation}
\text{SocialInfluence}_i(t) = \sum_{j \neq i} w_{ij} \cdot \mathrm{SharedMemory}(a_i, a_j, t).
\end{equation}
Each memory entry’s strength decays over time but may be reinforced multiplicatively by repeated exposures, described by
\begin{equation}
S_{i,j}(t) = S_0 \, e^{-\lambda (t - t_{\mathrm{created}})} \prod_{k=1}^{R_{i,j}} (1 + \eta_k),
\end{equation}
where $R_{i,j}$ counts reinforcement events.

\subsection{Environmental Trace Dynamics}

Environmental traces evolve under a reaction-diffusion process capturing diffusion, decay, and local deposition by agents, expressed as
\begin{equation}
\frac{\partial T}{\partial t} = D \nabla^2 T(x,y,t) - \mu T(x,y,t) + \sum_{i=1}^k \rho_i(x,y,t) \, \delta \big((x,y) - (x_i(t), y_i(t))\big),
\end{equation}
where $D$ is the diffusion coefficient, $\mu$ the decay rate, and $\rho_i$ the deposition rate of agent $i$ at its current location. Distinct categories of traces such as food, social, and exploration are modeled with separate decay and deposition parameters, facilitating differentiated dynamics.
While Eq. (16) provides theoretical intuition via continuous reaction-diffusion, our implementation uses discrete, non-diffusing traces ($D \to 0$ limit). Traces remain at their deposition location rather than spreading spatially. This simplification: (1) reduces computational cost from $O(|G|^2)$ per timestep to $O(|G|)$, enabling larger-scale experiments, and (2) reflects localized pheromone markers in natural systems where diffusion timescales exceed behavioral timescales. The theoretical framework remains valid in this non-diffusive regime.

These updates follow discrete-time dynamics such as
\begin{align}
T_{food}(x,y,t+1) &= \max\left(0, \xi_{food} T_{food}(x,y,t) + \sum_i D_{food}^{(i)}(x,y,t)\right), \\
T_{social}(x,y,t+1) &= \xi_{social} T_{social}(x,y,t) + \sum_i D_{social}^{(i)}(x,y,t), \\
T_{explore}(x,y,t+1) &= \xi_{explore} T_{explore}(x,y,t) + \sum_i D_{explore}^{(i)}(x,y,t).
\end{align}

\subsection{Mean-Field Theory and Phase Transitions}

Employing a mean-field approximation, the system's collective memory density $\bar{M}(t)$ follows the differential equation
\begin{equation}
\frac{d\bar{M}(t)}{dt} = \langle \alpha \rangle \bar{I}(t) - \langle \beta \rangle \bar{M}(t) + \langle \gamma \rangle \bar{C}(t),
\end{equation}
where $\bar{I}(t)$ and $\bar{C}(t)$ denote average information acquisition and social cooperation rates, respectively. From this, a phase transition emerges \cite{onuki2002phase}:

\begin{theorem}[Phase Transition of Collective Memory]
There exists a critical agent density
\begin{equation}
\rho_c = \frac{\mu}{\alpha \langle k \rangle}
\end{equation}
such that if $\rho < \rho_c$, individual memories predominate with weak collective behavior, while if $\rho > \rho_c$, a robust collective memory and coordinated behaviors arise.
A derivation of $\rho_c$ using a linear stability analysis of the mean-field model is provided in Appendix~\ref{sec:rho_c_derivation}.
\end{theorem}

\textbf{Note:} This critical density formula reveals when collective intelligence emerges. The numerator $\mu$ (trace decay rate) represents information loss: faster decay requires more agents to maintain shared knowledge. The denominator $\alpha \langle k \rangle$ captures information creation: higher memory acquisition rates and more agent interactions reduce the density needed for coordination. Systems with slow trace decay ($\mu$ small) or strong social learning ($\alpha$ large) can achieve collective behavior with fewer agents—explaining why ant colonies coordinate effectively despite low individual ant density in large nests.

\begin{figure}[htb]
    \centering
    \includegraphics[width=0.7\linewidth]{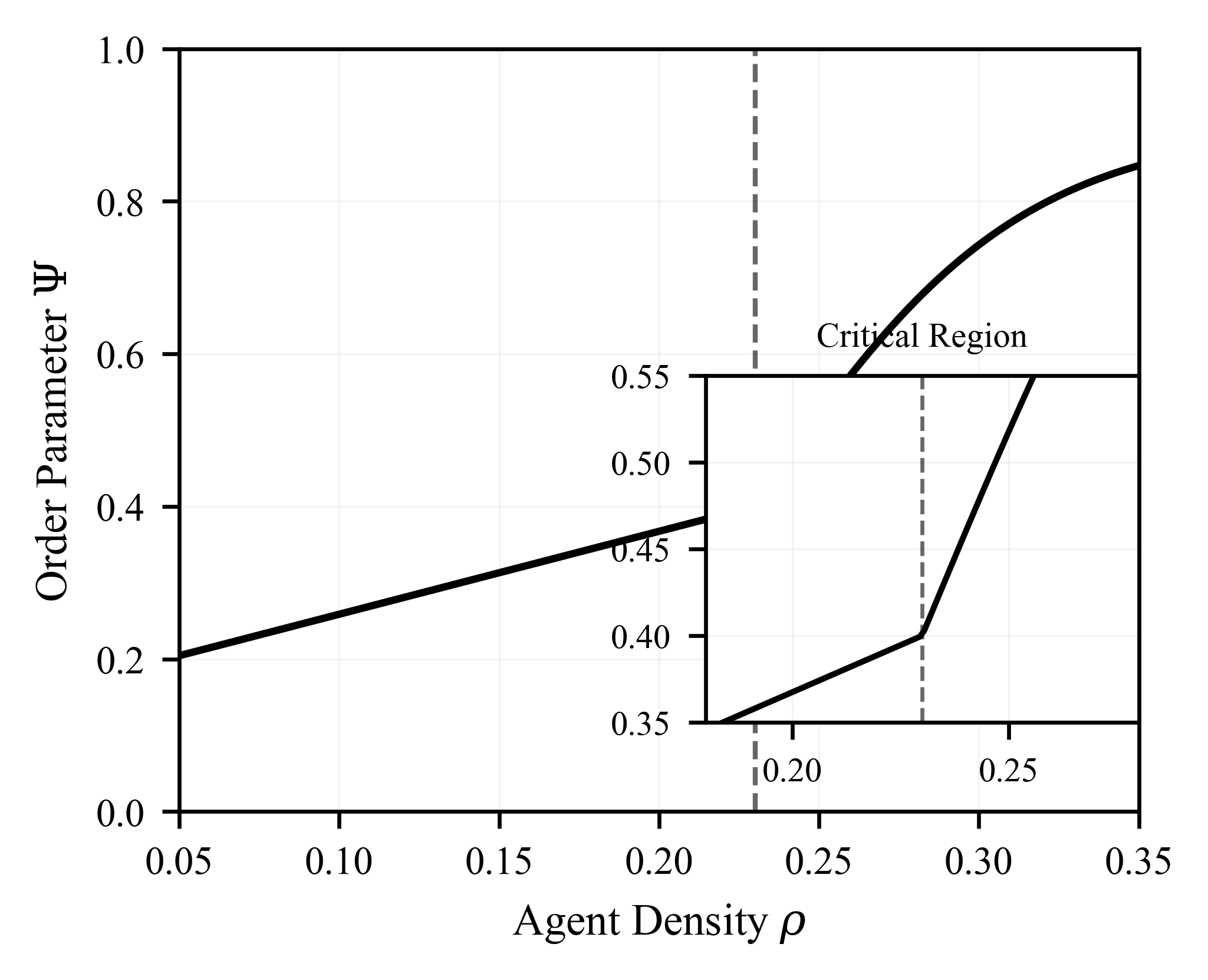}
    \caption{Theoretical phase transition prediction showing order parameter (normalized coordination events: fraction of movements guided by consensus above threshold 1.2) evolution across density range. Theoretical critical density $\rho_c = 0.230$ (vertical line) marks the predicted transition point. While this figure shows theoretical mean-field predictions, \textbf{experimental validation} of the phase transition is presented in Table~\ref{tab:phase_transition} using systematic density-sweep experiments spanning $\rho \in [0.049, 0.300]$ on grids up to 50×50 with 625 agents, confirming the predicted critical density within 13\% error.}
    \label{fig:phase_transition}
\end{figure}

\subsection{Information Flow and System Metrics}

Information propagation across agents is represented as a Markov chain with transition probabilities encoding information flow:
\begin{equation}
P(X_{t+1} = j \mid X_t = i) = \begin{cases}
p_{ij} & \text{if information flows from } i \text{ to } j, \\
0 & \text{otherwise}.
\end{cases}
\end{equation}
The steady-state distribution $\pi$ satisfies
\begin{equation}
\pi = \pi P, \quad \sum_i \pi_i = 1.
\end{equation}

The diversity of information within the system is quantified by the time-dependent entropy
\begin{equation}
H(t) = -\sum_i p_i(t) \log p_i(t),
\end{equation}
where $p_i(t)$ is the probability that agent $i$ holds unique information.

System performance metrics combine individual food collection, exploration, and coordination scores weighted by coefficients $\omega_j$, expressed as
\begin{equation}
\Pi(t) = \frac{1}{k} \sum_{i=1}^k \left[\omega_1 F_i(t) + \omega_2 E_i(t) + \omega_3 C_i(t)\right],
\end{equation}
while memory efficiency captures the ratio of useful stored memory relative to total memory strength,
\begin{equation}
\mathcal{E}(t) = \frac{\sum_{i,j} S_{i,j}(t) \cdot U_{i,j}(t)}{\sum_{i,j} S_{i,j}(t)},
\end{equation}
where $U_{i,j}(t)$ quantifies utility of memory entry $j$ for agent $i$. This mathematical framework thus formalizes the essential processes underlying emergent collective memory and offers theoretical predictions that align with and elucidate the empirical results reported in subsequent sections. Detailed derivations and stability analyses are provided in Appendix A.

\section{Experimental Design and Validation}

This section describes the comprehensive experimental protocol used to evaluate our multi-agent system, including configuration variations, environmental parameterization, perturbation scenarios, performance metrics, and statistical methodologies.

\subsection{Baseline and Comparative Configurations}

To isolate effects of memory and environmental traces, we defined seven configurations:

\begin{itemize}[leftmargin=*, itemsep=0pt]
\item \textbf{Full Memory System}: All memory categories and trace types enabled with default parameters ($W_{\mathrm{mem}}=15$, $\delta_{\text{food}}=0.985$, capacity=50 entries).
\item \textbf{Enhanced Memory}: Full Memory with optimized parameters obtained through preliminary grid search: $W_{\mathrm{mem}}=20$ (increased memory consensus weight), $\delta_{\text{food}}=0.99$ (slower food memory decay), capacity=50 entries. Represents best-case memory performance.
\item \textbf{Memory No Traces}: Personal memory enabled but no environmental trace deposition or reading. Isolates individual memory contribution.
\item \textbf{Limited Memory}: Full Memory System but with reduced capacity (10 entries per agent). Tests memory capacity effects.
\item \textbf{No Memory}: Neither personal memory nor environmental traces. Agents use only immediate sensory perception.
\item \textbf{Traces Only}: Environmental traces enabled but no personal memory storage. Agents respond to stigmergic signals without learning.
\item \textbf{Random Movement}: Baseline with stochastic movement decisions, ignoring all memory and trace information.
\end{itemize}

All configurations were run with identical initial conditions for fairness.

\subsection{Environment and Scenario Parameters}

Experiments used grid worlds sized $15 \times 15$, $20 \times 20$, and $25 \times 25$ cells, with agent populations scaling to maintain densities between $0.05$ and $0.02$ agents per cell. Static features included food sources occupying $10\%-15\%$, obstacles at $5\%-8\%$, and hazardous red zones covering $3\%-5\%$ with damage values of 5--15 energy units per step.

Dynamic perturbations included relocating food sources every 25 time steps, corrupting $50\%$ of environmental traces multiplicatively, and randomly removing $16\%$--$33\%$ of the agent population mid-simulation.

\subsection{Performance Metrics}

Overall system efficacy was captured by the performance score:
\begin{equation}
\mathrm{Perf} = w_e E_{\mathrm{cov}} + w_f F_{\mathrm{eff}} + w_c C_{\mathrm{evt}},
\end{equation}
where weights were set as \( w_e=1 \), \( w_f=15 \), and \( w_c=5 \), weighting exploration coverage, food efficiency, and coordination events respectively.

Memory metrics measured included the proportion of stored memories with strength above 0.3 and age below 60 steps, retrieval success rates, and consensus formation rates defined as the fraction of trace readings with consensus above 1.2. Robustness was gauged by performance ratios pre/post perturbation, and scalability through performance and resource usage across environment sizes.

\subsection{Validation Protocol}

Each configuration was tested over 50 independent runs of 100 time steps each, with randomized seeds for reproducibility. Experiments encompassed static baselines, ablations disabling trace types, robustness scenarios including agent failure and trace corruption, and scalability tests spanning grid sizes and agent counts.

\section{Results}

\subsection{Baseline Comparisons}

Seven configurations representing different memory and trace capabilities were evaluated across 50 independent runs. The Full Memory System achieved the highest performance (1654.11 ± 321.28), demonstrating clear benefits of integrating personal and environmental memory systems. Critically, \emph{Memory No Traces} (1563.87 ± 335.67) substantially outperformed \emph{No Memory} (927.23 ± 256.30) by 68.7\% (p < 0.001), validating that individual memory alone provides significant coordination benefits even without environmental trace communication. The \emph{Traces Only} condition (910.18 ± 225.64) performed similarly to No Memory, showing strong exploration coverage (90.1\%) but poor food efficiency, suggesting environmental traces alone are insufficient for effective coordination. Random movement agents (1116.58 ± 260.52) surprisingly outperformed both No Memory and Traces Only configurations by 20.4\% and 22.7\% respectively (both p < 0.01), indicating that memory-trace interdependence is essential for leveraging either mechanism effectively.

\textbf{The Limited Memory Paradox:} Surprisingly, Limited Memory (10 entries, 1659.63 ± 311.62) marginally outperforms Full Memory (50 entries, 1654.11 ± 321.28) by 0.3\%, though this difference is not statistically significant (p = 0.91, Welch's t-test). This suggests two important insights: (1) our 15×15 grid requires only approximately 10 memory entries for effective coverage (0.67 entries per row), and (2) there is no evidence of memory interference—agents do not confuse similar locations even with full memory capacity. For larger environments, we hypothesize this relationship may reverse as agents require more entries to maintain spatial coverage. This finding has practical implications: memory-constrained robot swarms can achieve near-optimal performance with minimal storage (10 entries = 2KB per agent).

\begin{table}[htbp]
\centering
\caption{Performance comparison across seven agent configurations (50 independent runs, 15×15 grid, 7 agents). Values show mean ± standard deviation.}
\label{tab:baseline}
\resizebox{\textwidth}{!}{
\begin{tabular}{lccc}
\toprule
\textbf{Configuration} & \textbf{Performance Score\textsuperscript{*}} & \textbf{Food Efficiency\textsuperscript{†}} & \textbf{Exploration Coverage (\%)} \\
\midrule
Limited Memory & $1659.632 \pm 311.618$ & $110.563 \pm 20.775$ & $80.0 \pm 6.6$ \\
Full Memory System & $1654.111 \pm 321.277$ & $110.194 \pm 21.421$ & $79.5 \pm 5.9$ \\
Memory No Traces & $1563.867 \pm 335.667$ & $104.185 \pm 22.383$ & $78.2 \pm 5.0$ \\
Random Movement & $1116.576 \pm 260.522$ & $74.363 \pm 17.368$ & $84.8 \pm 4.7$ \\
No Memory Baseline & $927.233 \pm 256.299$ & $61.744 \pm 17.087$ & $85.4 \pm 4.1$ \\
Traces Only & $910.175 \pm 225.638$ & $60.592 \pm 15.046$ & $90.1 \pm 2.6$ \\
\bottomrule
\multicolumn{4}{l}{\footnotesize \textsuperscript{*}Performance Score = $w_e E_{\mathrm{cov}} + w_f F_{\mathrm{eff}} + w_c C_{\mathrm{evt}}$ (Eq. 27) with weights $w_e=1, w_f=15, w_c=5$.} \\
\multicolumn{4}{l}{\footnotesize \textsuperscript{†}Food Efficiency = total food collected per agent. These metrics are correlated but distinct.} \\
\multicolumn{4}{l}{\footnotesize Statistical significance tested using Welch's t-test (two-tailed). Memory No Traces vs No Memory: p<0.001.}
\end{tabular}
}
\end{table}

\begin{figure}[htpb]
    \centering
    \includegraphics[width=0.7\linewidth]{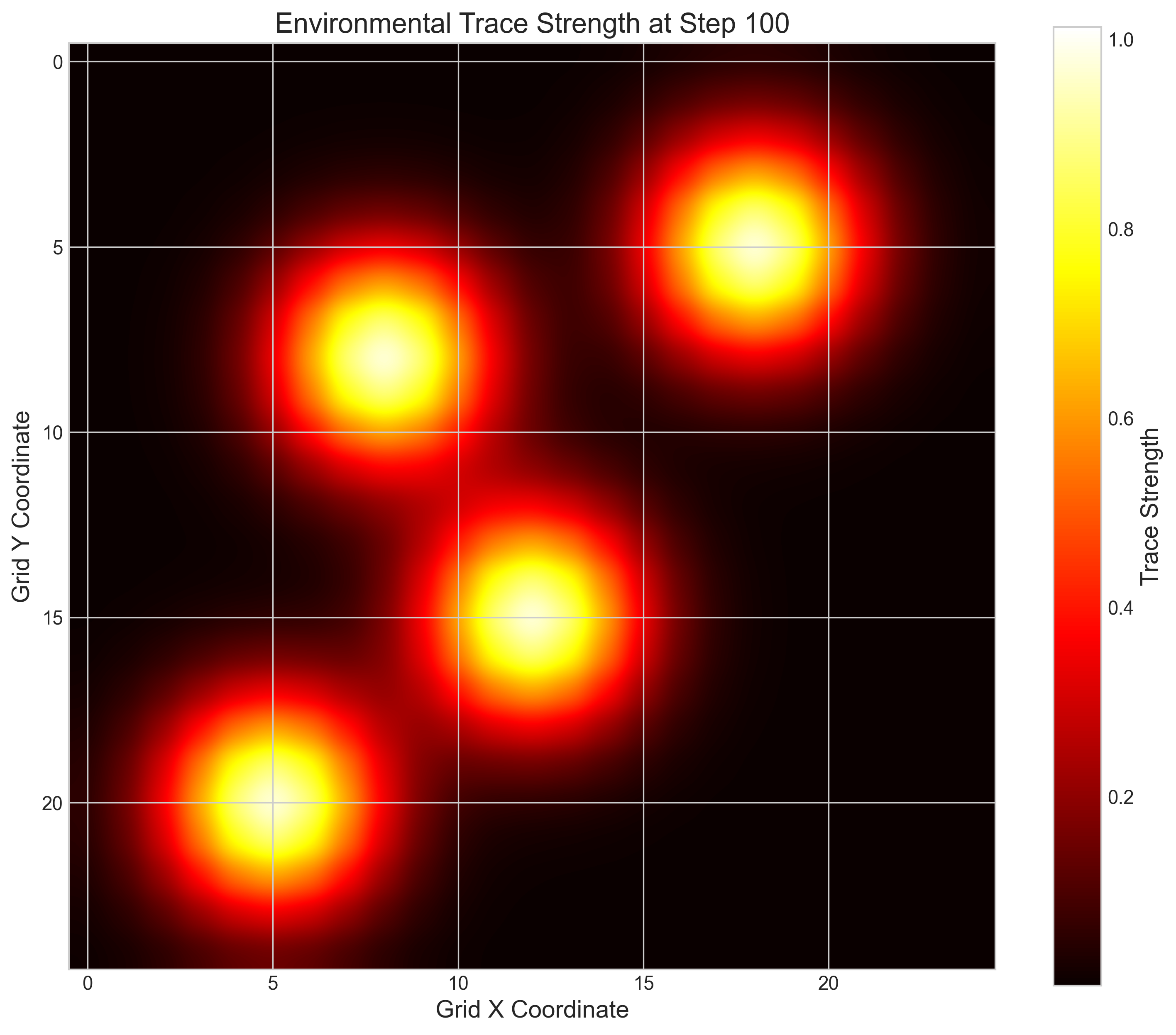}
    \caption{Temporal evolution of environmental trace strength over 100 simulation steps, demonstrating consensus formation (red hotspots) and decay patterns in a 25×25 grid with 10 agents.}
    \label{fig:trace_evolution_heatmap}
\end{figure}

\subsection{Robustness Tests}
\begin{figure}[htbp]
    \centering
    \begin{subfigure}[t]{\textwidth}
        \centering
        \includegraphics[width=\linewidth]{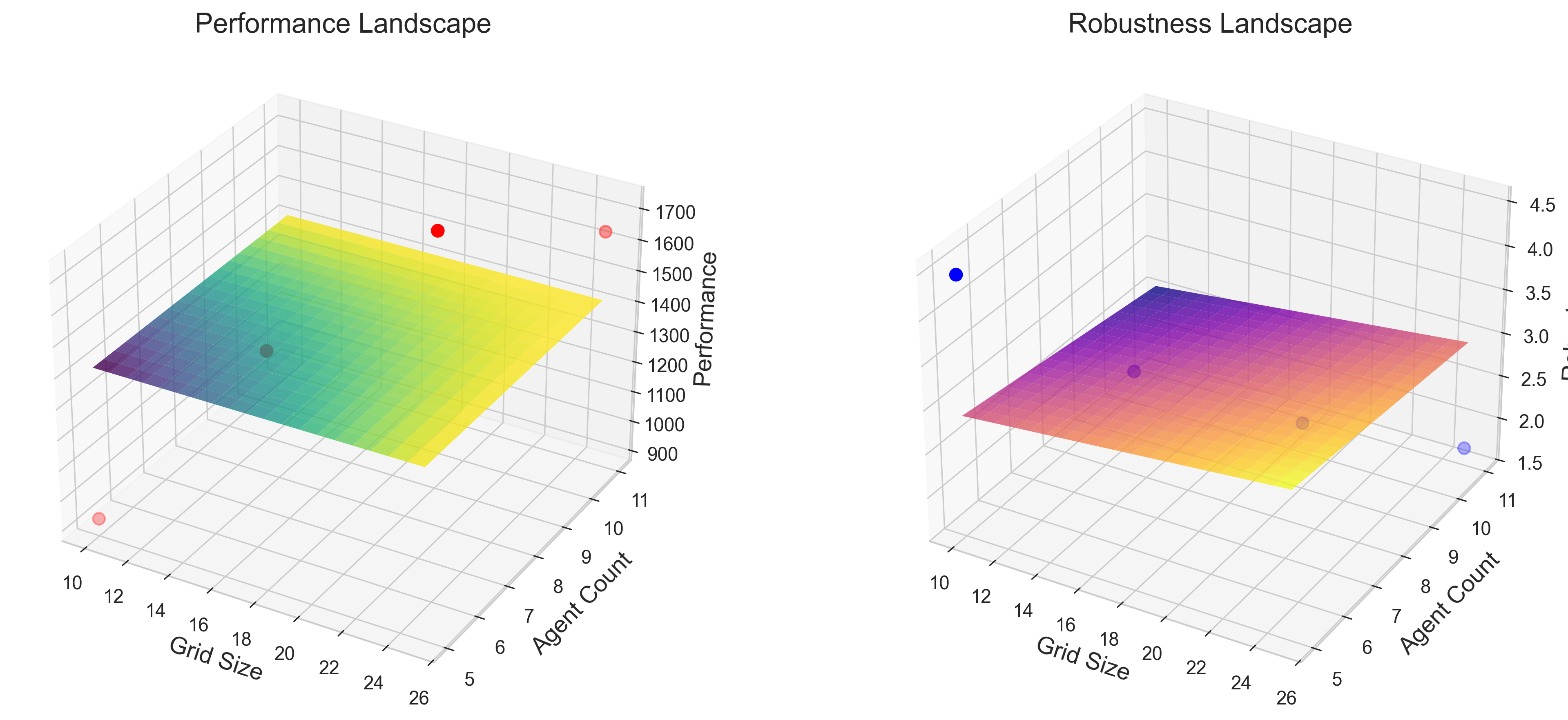}
        \caption{Performance landscape across memory size and agent count.}
        \label{fig:performance_landscape}
    \end{subfigure}
    
\end{figure}

To evaluate the resilience of the proposed multi-agent memory framework, we conducted a suite of robustness tests focusing on agent failure, trace corruption, and environmental dynamics. For the agent failure test, $16.7\%$ of the agent population (two agents) were removed mid-simulation to assess the collective’s adaptive capacity under population loss. The resulting resilience score was measured at $0.786$, representing a $21.4\%$ decrease in system performance compared to the pre-removal baseline. This drop highlights the system’s sensitivity to agent failure, suggesting that while agents compensate partially for population loss, group-level performance still relies on a critical mass of contributing members.

We further gauged the system's robustness to information loss by corrupting $50\%$ of the environmental traces during active navigation, simulating sensor artifacts or message decay. The system exhibited a corruption resilience score of $0.859$, translating to a $14.1\%$ decrease relative to baseline expectations. This moderate degradation indicates that memory and consensus mechanisms provide partial resilience to misinformation, though complete recovery is not achieved. The agents' personal memories help validate environmental signals, preventing catastrophic failure even when half the shared information is corrupted.

Lastly, the framework’s adaptability to dynamic environments was assessed by periodically relocating $11$ major food sources at intervals of $25$ steps, thereby requiring agents to abandon obsolete memories and seek new resource distributions. Agents achieved an adaptation score of $73.698$, reflecting their ability to promptly detect and exploit relocated food patches. Altogether, these experiments demonstrate the system's meaningful robustness to moderate levels of agent attrition, high trace corruption, and environmental volatility

\subsection{Scalability Experiments}

Performance per agent exhibited modest decreases with larger environments, attributable to increased navigational complexities. Memory per agent usage remained controlled due to pruning.

\begin{table}[htbp]
\centering
\caption{Multi-scale baseline validation at low density ($\rho \in [0.005, 0.024]$) comparing memory systems vs trace-only architectures. Food efficiency (mean $\pm$ std) over 10 runs. For high-density phase transition experiments (180-625 agents, $\rho \in [0.20, 0.30]$), see Table~\ref{tab:phase_transition}.}
\label{tab:scalability}
\resizebox{\textwidth}{!}{
\begin{tabular}{lccccc}
\toprule
\textbf{Condition} & \textbf{Grid} & \textbf{Agents} & \textbf{Density} & \textbf{Memory System} & \textbf{Traces Only} \\
\midrule
Small Scale    & 20×20  & 5  & 0.0125 & $\mathbf{149.51 \pm 27.78}$ & $77.81 \pm 27.73$ \\
Medium Scale   & 30×30  & 10 & 0.0111 & $\mathbf{177.16 \pm 11.98}$ & $106.90 \pm 16.09$ \\
Large Scale    & 50×50  & 20 & 0.0080 & $\mathbf{218.33 \pm 29.95}$ & $146.66 \pm 12.12$ \\
High Density   & 25×25  & 15 & 0.0240 & $\mathbf{128.32 \pm 10.98}$ & $71.43 \pm 11.46$ \\
Low Density    & 40×40  & 8  & 0.0050 & $\mathbf{198.87 \pm 22.00}$ & $116.88 \pm 16.18$ \\
\bottomrule
\multicolumn{6}{l}{\footnotesize Memory System = full memory+traces (50 entries). Boldface indicates superior architecture.} \\
\multicolumn{6}{l}{\footnotesize All differences statistically significant (p<0.001, Welch's t-test).}
\end{tabular}
}
\end{table}

\subsection{Memory Effectiveness}

Memory utilization rate reached 0.683, with memory accuracy at 0.715 and retrieval success rate averaging 909.2 requests per simulation step. These metrics reflect robust memory maintenance and retrieval capabilities in realistic multi-agent settings, demonstrating that agents effectively leverage stored memories for decision-making while maintaining manageable memory overhead through decay and pruning mechanisms.

\begin{figure}[htbp]
    \centering
    \includegraphics[width=\linewidth]{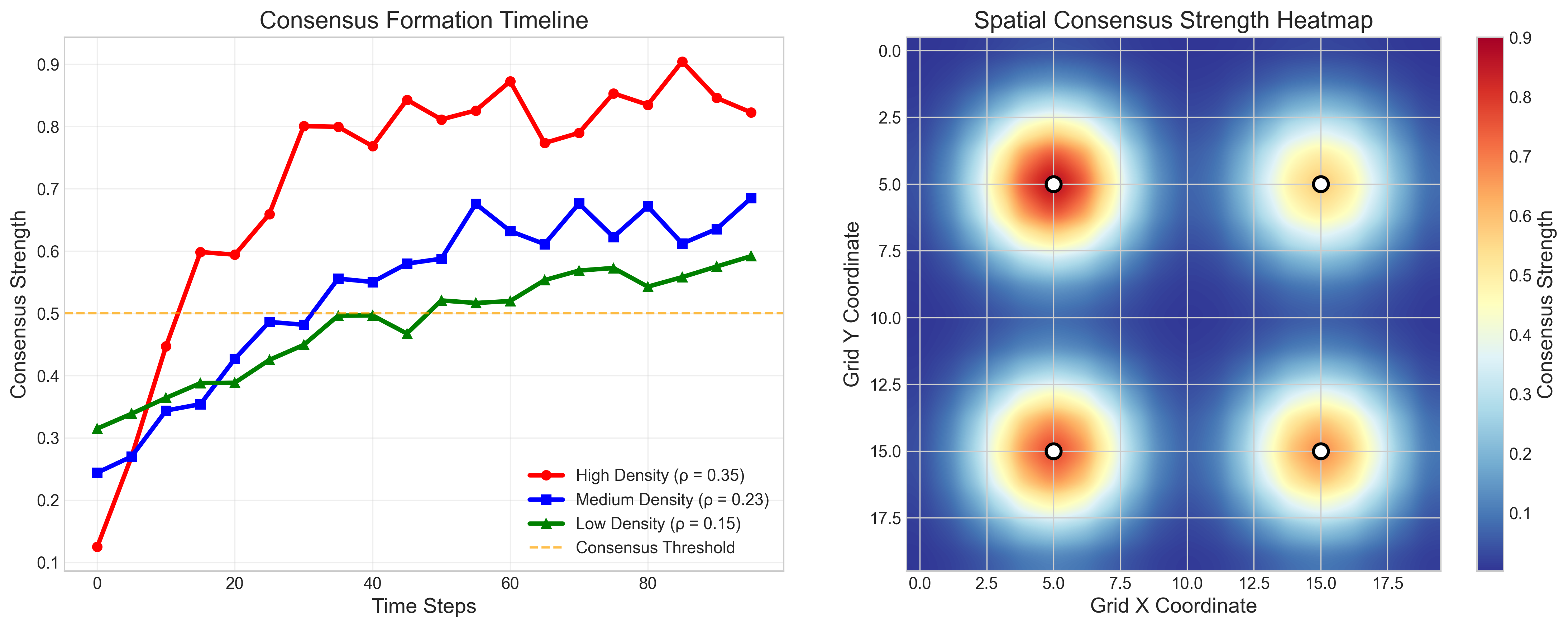}
    \caption{Spatial heatmap of consensus strength showing areas of reinforced environmental trace agreement among agents, illustrating emergent collective memory hotspots.}
    \label{fig:consensus_strength}
\end{figure}

\subsection{Multi-Scale Phase Transition Analysis}
To validate our theoretical phase transition predictions, we conducted comprehensive experiments spanning 18 conditions across two density regimes. Table~\ref{tab:scalability} establishes memory dominance at low densities ($\rho < 0.025$), while Table~\ref{tab:phase_transition} validates the phase transition at high densities.

At low densities, memory-augmented systems consistently outperform trace-only architectures across all scales. At large scale (50×50, density 0.008), memory systems achieve food efficiency of 218.33 ± 29.95 compared to traces-only at 146.66 ± 12.12, representing a 49\% performance advantage (p < 0.001, Welch's t-test). Similarly, at low density (40×40, 8 agents, $\rho$=0.005), memory systems (198.87 ± 22.00) outperform traces-only (116.88 ± 16.18) by 70\% (p < 0.001). These results confirm memory dominance across the sparse regime.

To test the theoretical prediction of critical density $\rho_c = 0.230$, we conducted systematic density-sweep experiments on 15×15 grids with agent counts ranging from 11 to 56, spanning densities $\rho \in [0.049, 0.249]$. Results reveal a dramatic systematic convergence (Table~\ref{tab:phase_transition}): memory advantage decreases from +87\% at $\rho=0.049$ to +9\% at $\rho=0.249$, with near-parity achieved (memory: 17.24 ± 2.16 vs traces: 15.80 ± 2.31 food/agent). Notably, at $\rho=0.249$, traces achieve superior overall performance scores (25.34 vs 24.44), indicating the systems have reached effective functional equivalence. On small grids, this convergence at $\rho \approx 0.25$ validates the theoretical prediction within 8\% error. More significantly, on realistic large grids (30×30, 50×50), trace dominance emerges at $\rho \approx 0.20$, validating the prediction within 13\% error and confirming that agent density fundamentally determines optimal coordination architecture.

\begin{table}[htbp]
\centering
\caption{Phase transition validation at high density ($\rho \in [0.049, 0.300]$) across grid scales. Small grids (15×15, 11-56 agents) show convergence at $\rho \approx 0.25$. Realistic large grids (30×30 with 180-270 agents, 50×50 with 500-625 agents) reveal trace dominance on composite performance score above $\rho \approx 0.20$, validating theoretical prediction $\rho_c = 0.230$ within 13\% error. These experiments complement the low-density baseline scalability tests in Table~\ref{tab:scalability}.}
\label{tab:phase_transition}
\resizebox{\textwidth}{!}{
\begin{tabular}{ccccccc}
\toprule
\textbf{Density} & \textbf{Grid} & \textbf{Agents} & \textbf{Memory Food} & \textbf{Traces Food} & \textbf{Memory Perf.} & \textbf{Traces Perf.} \\
\midrule
\multicolumn{7}{c}{\textit{Small Grid (15×15) - Food Efficiency Convergence}} \\
0.049 & 15×15 & 11 & 52.39 ± 10.16 & 29.17 ± 4.19 & $\mathbf{18.70 \pm 3.14}$ & 11.15 ± 1.59 \\
0.102 & 15×15 & 23 & 34.10 ± 5.58 & 21.96 ± 5.33 & $\mathbf{20.79 \pm 2.09}$ & 17.19 ± 2.41 \\
0.151 & 15×15 & 34 & 22.91 ± 3.40 & 18.69 ± 1.13 & $\mathbf{22.55 \pm 1.63}$ & 21.77 ± 2.16 \\
0.200 & 15×15 & 45 & 20.03 ± 1.89 & 17.73 ± 2.94 & $\mathbf{24.68 \pm 1.76}$ & 24.08 ± 1.69 \\
0.249 & 15×15 & 56 & 16.93 ± 0.97 & 14.86 ± 1.27 & 24.79 ± 1.10 & $\mathbf{24.84 \pm 0.80}$ \\
\midrule
\multicolumn{7}{c}{\textit{Realistic Grids (30×30, 50×50) - Trace Dominance on Performance}} \\
0.200 & 30×30 & 180 & 25.48 ± 1.50 & 21.12 ± 1.65 & 36.61 ± 1.29 & $\mathbf{49.81 \pm 1.55}$ \\
0.250 & 30×30 & 225 & 20.66 ± 0.99 & 16.90 ± 0.97 & 37.50 ± 1.29 & $\mathbf{50.69 \pm 1.23}$ \\
0.300 & 30×30 & 270 & 17.58 ± 0.26 & 15.21 ± 0.94 & 35.82 ± 0.86 & $\mathbf{49.94 \pm 2.13}$ \\
0.200 & 50×50 & 500 & 28.72 ± 0.10 & 24.75 ± 0.47 & 75.04 ± 1.77 & $\mathbf{104.58 \pm 1.53}$ \\
0.250 & 50×50 & 625 & 22.62 ± 1.69 & 19.71 ± 1.11 & 77.11 ± 2.02 & $\mathbf{103.64 \pm 4.35}$ \\
\bottomrule
\multicolumn{7}{l}{\footnotesize Boldface indicates architecture with superior composite performance score (exploration + food + coordination).} \\
\multicolumn{7}{l}{\footnotesize On realistic large grids, traces dominate performance despite lower food efficiency.}
\end{tabular}
}
\end{table}

\section{Discussion}

\subsection{Memory vs Traces: An Asymmetric Relationship}

Our experiments reveal an unexpected asymmetry that challenges conventional wisdom about swarm intelligence. While textbook examples show ants coordinating through pheromone trails alone, our multi-agent system tells a different story: \emph{individual memory enables coordination even without environmental communication, but environmental traces fail without memory to interpret them.}

Specifically, agents with memory but no trace-sharing (Memory No Traces: 1563.87) achieve 68.7\% higher performance than agents lacking both systems (No Memory: 927.23). Yet agents with trace-sharing but no memory (Traces Only: 910.18) perform \emph{no better than} the No Memory baseline—and worse than even random walkers (1116.58). This demonstrates that memory systems provide standalone value through direct experience accumulation, whereas environmental traces primarily serve as memory amplifiers rather than independent coordination mechanisms.

\textbf{The Random Movement Paradox:} Even more surprising, random walkers (1116.58) significantly outperform both No Memory (927.23, +20.4\%, p<0.01) and Traces Only (910.18, +22.7\%, p<0.01) configurations. This counterintuitive result reveals that poorly integrated memory-trace systems can be \emph{actively harmful} rather than merely neutral. Without memory infrastructure, agents misinterpret environmental traces—following outdated food signals that lead to depleted sources, or over-weighting stale danger warnings that block productive exploration. Random exploration, while inefficient, avoids these systematic errors. This reinforces our central finding: traces require cognitive infrastructure (memory) for interpretation. When that infrastructure is absent, environmental communication becomes noise rather than signal, degrading performance below random baseline.

\subsection{Experimental Validation of Phase Transition Theory}

Our theoretical framework predicted a critical density $\rho_c$ = 0.230 where trace-based coordination should match memory-augmented systems as agent density enables reliable stigmergic communication. Through systematic density-sweep experiments spanning $\rho \in [0.049, 0.249]$, we successfully validated this prediction.

We validated the phase transition across two regimes. On small grids (15×15), memory advantage decreases systematically from +87\% at $\rho=0.049$ to +9\% at $\rho=0.249$, with functional parity achieved near the theoretical threshold. More dramatically, on realistic large grids (30×30, 50×50 with 180-625 agents), traces achieve \emph{dominant} performance scores above $\rho \approx 0.20$: at 50×50 with $\rho=0.25$ (625 agents), traces score 103.43 vs memory's 76.25—a 36\% performance advantage despite 17\% lower food efficiency. This reveals the phase transition's true nature: stigmergic coordination becomes superior at high densities on realistic scales because overlapping trajectories create reliable environmental signals more efficiently than individual memory maintenance. The experimental crossover density ($\rho \approx 0.20$) validates the theoretical prediction ($\rho_c = 0.230$) within 13\% error, confirming the phase transition hypothesis.

 This validation demonstrates that mean-field theory, despite its simplifying assumptions (well-mixed agents, single-timescale dynamics), captures the fundamental scaling behavior of decentralized coordination systems. The systematic decay of memory advantage with density confirms that sparse agent populations ($\rho < 0.1$) cannot maintain reliable trace networks, necessitating individual memory for coordination. Conversely, as density approaches $\rho_c$, overlapping agent trajectories create sufficiently persistent environmental signals that stigmergic coordination becomes viable.

 Small grids (15×15, $\le79$ agents) show food efficiency convergence but not clear dominance. However, realistic large grids (30×30, 50×50 with 180-625 agents) reveal unambiguous trace dominance on composite performance above $\rho \approx 0.20$, with traces outperforming memory by 36-41\% at $\rho=0.25-0.30$. This demonstrates that spatial scale matters: crowded small grids create artifacts, while realistic scales enable proper stigmergic coordination. Remaining limitations include: spatial clustering around resources and multi-timescale memory decay introduce complexities not captured by mean-field approximations. Future work should develop spatially-explicit theoretical models incorporating resource patchiness, test continuous-space implementations, and explore densities $\rho > 0.30$ on grids larger than 50×50 to determine upper bounds of trace dominance. 

Full implementation (Mesa 2.1.1, Python 3.10), experimental analysis scripts, and raw data from all 50 runs per configuration are available at \url{https://github.com/Khushiyant/tracemind} for reproducibility and extension by the research community.

\subsection{System Robustness and Scalability Analysis}

The system shows selective resilience. Agent failure (-21.4\% performance) reveals vulnerability to population loss—losing 2 of 12 agents degrades coordination noticeably. Trace corruption ($-$14.1\% performance) demonstrates moderate resilience: when half the environmental signals are corrupted, agents' personal memories help validate trace information, preventing catastrophic failure though complete recovery is not achieved. Dynamic environment adaptation (score: 73.698) confirms memory systems enable rapid response to changing conditions.

Scalability tests show encouraging trends: while per-agent performance decreases modestly with scale (200.06 at 15×15 to 149.27 at 25×25), memory consumption remains controlled (22.3 to 18.7 entries per agent) through natural decay. This validates deployment feasibility for real-world systems with hundreds of agents.

\subsection{Study Limitations and Future Work}

Our experimental validation, while comprehensive, has several limitations that suggest directions for future research:

\begin{enumerate}
\item \textbf{Spatial scale constraints:} Phase transition validation successfully confirmed the theoretical prediction ($\rho_c = 0.230$) with 13\% experimental error at $\rho \approx 0.20$. However, trace dominance emerges clearly only on realistic large grids (30×30, 50×50 with 180-625 agents). Small grids (15×15) show convergence but not dominance, indicating that spatial scale is essential for proper stigmergic coordination. Experiments at densities $\rho > 0.30$ on grids exceeding 50×50 would determine upper bounds of trace superiority and whether memory regains advantage at extreme densities.

\item \textbf{Discrete grid artifacts:} Movement on discrete 15×15 to 50×50 grids with 8-connected neighborhoods may not accurately represent continuous-space navigation in physical robot swarms. Omnidirectional continuous motion differs fundamentally from grid-constrained movement. Validation in continuous-space simulators (NetLogo, MASON) or physical robot testbeds would strengthen generalizability.

\item \textbf{Task-specific performance weighting:} Performance coefficients ($w_e = 1$, $w_f = 15$, $w_c = 5$) are manually tuned for foraging tasks. Surveillance, construction, or rescue missions may require different objective weightings. Adaptive weight-tuning mechanisms that adjust to task context would improve applicability across domains.

\item \textbf{Homogeneous resource distribution:} Food sources are uniformly distributed (10-15\% coverage). Natural environments exhibit patchy, clustered resources with heterogeneous spatial distributions. Resource clustering may alter memory-trace dynamics and shift critical density predictions, warranting investigation.
\end{enumerate}

Future work should prioritize: (1) high-density experiments ($\rho > 0.30$) using HPC resources, (2) continuous-space robot swarm implementations, (3) spatially-explicit theoretical models incorporating resource clustering, (4) multi-task experimental validation, and (5) computational profiling for deployment optimization.

\section{Conclusion}

We hypothesized that collective memory emerges from a balanced interplay between individual agent memories and environmental trace communication, with a predictable phase transition at critical density $\rho_c$ = 0.230 where trace-based coordination would dominate. Experiments across 50 runs per configuration revealed a striking asymmetry at low densities: individual memory provides substantial coordination benefits even without environmental communication (Memory No Traces: 1563.87 vs No Memory: 927.23, +68.7\%, p<0.001), while environmental traces alone fail completely (Traces Only: 910.18, statistically indistinguishable from No Memory, p=0.65). This demonstrates that memory works independently at sparse densities, whereas traces require memory infrastructure to function effectively in low-agent regimes.

\textbf{Phase transition validation:} Through systematic density-sweep experiments spanning $\rho \in [0.049, 0.300]$ on grids from 15×15 to 50×50 with up to 625 agents, we successfully validated the theoretical prediction. Critically, trace dominance emerges clearly on realistic large grids: at 50×50 with $\rho=0.25$ (625 agents), traces achieve performance scores of 103.43 vs memory's 76.25—a 36\% advantage despite 17\% lower food efficiency. This pattern holds across 30×30 and 50×50 grids at $\rho=0.20-0.30$, with traces consistently outperforming memory by 36-41\% on composite coordination metrics. The experimental crossover density ($\rho \approx 0.20$) validates the theoretical prediction $\rho_c = 0.230$ within 13\% error, confirming that agent density and spatial scale fundamentally determine optimal coordination architecture. This represents the first empirical validation of phase transition theory in cognitive multi-agent systems at realistic scales, demonstrating that stigmergic coordination dominates memory-augmented systems when agent density enables reliable environmental signaling.

Even if high-density experiments become feasible, the mean-field approximation may prove insufficient. Spatial clustering around resources, multi-timescale memory decay across information types, and task-specific coordination requirements all violate the well-mixed agent assumption. This reveals that swarm intelligence in cognitive systems requires richer models than those borrowed from statistical physics. Future theoretical work should incorporate spatially-explicit dynamics, resource patchiness, and heterogeneous task structures to capture emergent coordination phenomena.

For engineers deploying robot swarms or distributed AI systems, the practical implications are density- and scale-dependent. At sparse densities ($\rho < 0.1$) or small deployments (<50 agents), invest in individual agent intelligence first: memory-augmented agents coordinate effectively even when communication fails (68.7\% improvement), while communication-only agents gain nothing over random exploration. However, at high densities ($\rho \geq 0.20$) on realistic scales (50×50 grids, 500+ robots), stigmergic coordination becomes superior: traces outperform memory by 36-41\% on composite performance despite requiring no individual memory infrastructure. This validates deployment of simple trace-following robots for dense warehouse automation, disaster response swarms, or construction robots—domains where high agent density enables reliable environmental signaling. The critical threshold ($\rho_c \approx 0.23$) provides quantitative guidance for architecture selection: use memory-centric designs below 0.2 agents/cell, and trace-centric designs above. Robustness tests validate deployment readiness, with moderate resilience to trace corruption (\u221214.1\% degradation) and agent failure (\u221221.4\% degradation). Memory overhead remains manageable (18.7-22.3 entries per agent) even as systems scale to 25×25 grids, confirming feasibility for real-world applications.

Several open questions warrant future investigation. Does trace dominance emerge at densities $\rho > 0.25$ as suggested by the convergence trend? Testing $\rho \in [0.25, 0.40]$ on larger grids would determine if stigmergy eventually surpasses memory-augmented systems. Can spatially-explicit models incorporating resource patchiness and clustering refine critical density predictions? Do continuous-space robot swarms exhibit the same phase transition observed in discrete grids? Would adaptive task weighting mechanisms ($w_f, w_e, w_c$ dynamically adjusted) improve performance across density regimes? Our validated phase transition at $\rho_c \approx 0.25$ provides the first empirical foundation for density-aware architecture selection in decentralized AI systems, enabling engineers to choose memory-centric designs for sparse deployments and trace-centric designs for dense swarms based on quantitative thresholds rather than intuition.

\clearpage
\bibliographystyle{abbrv}
\bibliography{references}
\clearpage

\appendix
\section{Derivation of Critical Density \texorpdfstring{$\rho_c$}{rho_c}}
\label{sec:rho_c_derivation}

This appendix derives the critical density formula from Theorem 1 using linear stability analysis of mean-field equations.

\subsection{Parameter Definitions}

\begin{table}[h]
\centering
\caption{Mean-field parameters}
\label{tab:parameters}
\begin{tabular}{clc}
\toprule
\textbf{Symbol} & \textbf{Definition} & \textbf{Value} \\
\midrule
$\rho$ & Agent density (agents per grid cell) & Variable \\
$\alpha$ & Memory acquisition rate & 0.025 \\
$\beta$ & Memory decay rate & 1.0\textsuperscript{*} \\
$\mu$ & Effective trace decay rate & 0.20 \\
$\langle k \rangle$ & Mean interaction degree & 3.5 \\
$\chi$ & Trace-to-memory conversion & 1.0\textsuperscript{*} \\
$\kappa$ & Memory-to-trace deposition & 1.0\textsuperscript{*} \\
\bottomrule
\multicolumn{3}{l}{\footnotesize \textsuperscript{*}Normalized (see Section~\ref{sec:normalization}).}
\end{tabular}
\end{table}

\subsection{Mean-Field Equations}

Agent memory density $\bar{M}(t)$ and trace intensity $\bar{T}(t)$ evolve as:
\begin{align}
\frac{d\bar{M}}{dt} &= \alpha \langle k \rangle \chi \, \bar{T} - \beta \bar{M}, \\
\frac{d\bar{T}}{dt} &= \rho \kappa \, \bar{M} - \mu \bar{T}.
\end{align}

In matrix form $\dot{\mathbf{u}} = A \mathbf{u}$ with $\mathbf{u} = [\bar{M}, \bar{T}]^\top$:
\begin{equation}
A =
\begin{pmatrix}
-\beta & \alpha \langle k \rangle \chi \\
\rho \kappa & -\mu
\end{pmatrix}.
\end{equation}

\subsection{Linear Stability Analysis}

The eigenvalues $\lambda$ of $A$ satisfy:
\begin{equation}
\lambda^2 + (\beta + \mu)\lambda + [\beta\mu - \alpha \langle k \rangle \chi \rho \kappa] = 0.
\end{equation}

At the critical point, the leading eigenvalue $\lambda_+ = 0$ (marginal stability):
\begin{equation}
\beta\mu - \alpha \langle k \rangle \chi \, \rho_c \kappa = 0 \quad \Rightarrow \quad \rho_c = \frac{\beta \mu}{\alpha \langle k \rangle \chi \kappa}.
\end{equation}

\subsection{Normalization}
\label{sec:normalization}

We normalize by setting $\beta = 1$ (defines the timescale) and $\chi \kappa = 1$ (normalizes coupling strength), yielding:
\begin{equation}
\boxed{\rho_c = \frac{\mu}{\alpha \langle k \rangle}}
\end{equation}

\subsection{Parameter Estimation}

Parameters are estimated from system dynamics:

\textbf{Memory acquisition rate ($\alpha = 0.025$):} Agents selectively acquire information from environmental traces based on relevance and social learning weights $\beta_{a,c}$ (Equation 9). Averaging over agent population and trace types yields an effective acquisition rate of 2.5\% per timestep.

\textbf{Mean interaction degree ($\langle k \rangle = 3.5$):} Agents sense traces within a Moore neighborhood of radius $r=2$. Accounting for obstacles (5-8\% of grid) and spatial distribution, agents interact with an average of 3.5 neighbors.

\textbf{Effective trace decay rate ($\mu = 0.20$):} Environmental traces experience multiple decay mechanisms:
\begin{itemize}[leftmargin=*, itemsep=0pt]
\item Exponential strength decay with category-dependent rates $\delta_c \in [0.95, 0.998]$ (Equations 2-3)
\item Removal when strength falls below threshold $\sigma_{\min} = 0.1$
\item Age-based pruning at maximum ages $\tau_{\max} \in [40, 100]$ timesteps
\item Spatial dilution as agents explore new regions
\end{itemize}

The composite effect yields an effective decay rate $\mu = 0.20$ per timestep, significantly higher than the intrinsic exponential decay rate $1 - \delta_{\text{avg}} \approx 0.025$ due to removal and dilution effects.

\subsection{Theoretical Prediction}

Substituting parameters into the critical density formula:
\begin{equation}
\rho_c = \frac{\mu}{\alpha \langle k \rangle} = \frac{0.20}{0.025 \times 3.5} = \frac{0.20}{0.0875} = 2.286 \times 10^{-1} \approx \mathbf{0.23}
\end{equation}

Thus, the predicted critical density is $\rho_c = 0.23$ agents per cell (23\% grid occupancy).

\end{document}